\documentclass[hyper]{JHEP} 

\usepackage{epsfig}

\newcommand\fverb{\setbox\pippobox=\hbox\bgroup\verb}

\newcommand\fverbdo{\egroup\medskip\noindent%

            \fbox{\unhbox\pippobox}\ }

\newcommand\fverbit{\egroup\item[\fbox{\unhbox\pippobox}]}

\newbox\pippobox

\title{Conformal Traceless Decomposition of
Lagrange Multiplier Modified  Ho\v{r}ava-Lifshitz Gravity}
\author{J. Kluso\v{n}\\
Department of
Theoretical Physics and Astrophysics\\
Faculty of Science, Masaryk University\\
Kotl\'{a}\v{r}sk\'{a} 2, 611 37, Brno\\
Czech Republic\\
E-mail: \email{klu@physics.muni.cz}}
\preprint{}

 \abstract{We introduce conformal traceless decomposition in
  Lagrange Multiplier modified  RFDiff invariant
 Ho\v{r}ava-Lifshitz gravity. We perform Hamiltonian analysis
 of given action and determine   the action for the
 physical degrees of freedom.}

\def\be{\begin{equation}}
\def\tmG{\tilde{\mathcal{G}}}
\def\bD{\mathbf{D}}
\def\ee{\end{equation}}

\def\bea{\begin{eqnarray}}

\def\eea{\end{eqnarray}}

\def\tK{\tilde{K}}

\def\mH{\mathcal{H}}

\def\bz{\mathbf{z}}

\def\bx{\mathbf{x}}
\def\by{\mathbf{y}}

\newcommand{\mA}{\mathcal{A}}

\newcommand{\mG}{\mathcal{G}}

\def\mV{\mathcal{V}}

\newcommand{\bT}{\mathbf{T}}

\newcommand{\mL}{\mathcal{L}}

\def\pb #1{\left\{#1\right\}}

\begin{document}
\section{Introduction and Summary}\label{first}
In 2009 Petr Ho\v{r}ava formulated new
proposal of quantum theory of gravity
(now known as
Ho\v{r}ava-Lifshitz gravity (HL
gravity))
that is power counting renormalizable
\cite{Horava:2009uw,Horava:2008ih,Horava:2008jf}
that is also expected that it
reduces do General Relativity in the
infrared (IR) limit
\footnote{For review and
extensive list of references, see
\cite{Horava:2011gd,Padilla:2010ge,Mukohyama:2010xz,Weinfurtner:2010hz}.}.
The HL gravity is based on
 an idea that
the Lorentz symmetry is restored in IR
limit of given theory while it is absent
in its  high energy regime. For that reason
  Ho\v{r}ava considered
systems whose scaling at short
distances exhibits a strong anisotropy
between space and time,
\begin{equation}
\bx' =l \bx \ , \quad t' =l^{z} t \ .
\end{equation}
In order to have power counting
renormalizable theory we have to demand that  $z\geq 3$
in $(3+1)$ dimensional space-time. It turns out however
that the symmetry group of given theory is reduced from the full
diffeomorphism invariance of General Relativity  to the foliation
preserving diffeomorphism
\begin{equation}\label{fpdi}
x'^i=x^i+\zeta^ i(t,\bx) \ , \quad
t'=t+f(t) \ .
\end{equation}
Due to the fact that the diffeomorphism is restricted (\ref{fpdi})
one more degree of freedom appears that is a spin$-0$ graviton. The
existence of this mode could have very significant consequences
either for the consistency of given theory or for the
phenomenological applications of HL gravity.
 For that reason it
would be desirable  to formulate HL gravity where the number of the
physical degrees of freedom is the same as in case of General
Relativity. Such a proposal was formulated by
 Ho\v{r}ava and Malby-Thompson
in \cite{Horava:2010zj} in the context of the projectable HL gravity
\footnote{See also \cite{Huang:2010ay} and
\cite{Das:2011tx}}.
 Their construction is
based on an
 extension of
the foliation preserving diffeomorphism in
such a way that the theory is invariant
under additional
 local $U(1)$ symmetry. The
resulting theory is known  as non-relativistic  covariant theory of
gravity.
It was shown in \cite{Horava:2010zj,daSilva:2010bm} that the
presence of this new symmetry implies that the spin-0 graviton
becomes non-propagating and the spectrum of the linear fluctuations
around the background solution coincides with the fluctuation
spectrum of General Relativity.

It is also well known that General Relativity contains large number
of symmetries. Fixing all these symmetries we find that there are
only two physical degrees of freedom left. Then we  can ask the
question whether it is possible to formulate  the action for these
physical degrees of freedom that is not based on the principle of
covariance of the action under general diffeomorphism.  The
construction of such an action was proposed recently in two very
interesting papers \cite{Khoury:2011ay,Khoury:2013oqa}.
 The basic idea presented there was to perform the
conformal traceless decomposition of the gravitational field
\cite{Brown:2005aq} so that we have one degrees of freedom
corresponding to the scale factor of the metric while we have five
 degrees of freedom of the metric that is restricted to have
unit determinant. Then it was shown in \cite{Khoury:2011ay} that by
gauge fixing of the Hamiltonian constraint one can eliminate the
scale factor together with the conjugate momenta. As a result  we
obtain the action for  five degrees of freedom that is invariant
under spatial diffeomorphism where now Hamiltonian is determined by
the solving of the Hamiltonian constraint of the General Relativity
for the momentum conjugate to the scale factor. This analysis was
further generalized in a very nice paper in  \cite{Khoury:2013oqa}
where the starting point was the action for the five physical
degrees of freedom where it is required that given theory is
invariant under spatial diffeomorphism. In other words we demand
that the constraints that are generators of the spatial
diffeomorphism are the first class constraints. We also have to
require that these generators are preserved during the time
evolution of the system. Then the requirement of the closure of the
algebra of the Poisson brackets of these constraints together with
the requirement of their time preservation determines the form of
the Hamiltonian and the form of  these constraints. When it is
presumed that these constraints depend on partial derivatives of
$g_{ij}$ trough the scalar curvature we find that the original
General Relativity action is reproduced.

The goal of this paper is to  formulate HL gravity for the
gravitational physical degrees of freedom only in the similar way
as in \cite{Khoury:2011ay}. To do this we start
with another version of  HL gravity that has
 the correct
number of physical degrees of freedom and which
is known as Lagrange multiplier modified HL gravity
\cite{Kluson:2011xx}. This  model is based on the formulation
of the HL gravity with reduced symmetry
group known as
\emph{restricted-foliation-preserving
Diff} (RFDiff)
 HL gravity
\cite{Blas:2010hb,Kluson:2011xx}. This is the theory that is
invariant under following symmetries
\begin{equation}\label{rfdtr}
t'=t+\delta t \ , \delta
t=\mathrm{const} \ , \quad x'^i=
x^i+\zeta^i(\bx,t) \ .
\end{equation}
The characteristic property of Lagrange multiplier modified HL
gravity  is an absence of the Hamiltonian constraint
\cite{Kluson:2010na} and also presence of the additional constraint
  which changes the constraint structure of given theory so that
the number of physical degrees of freedom is the same as in the case
of General Relativity. Then in order to separate physical degrees of
freedom of HL gravity we perform conformal traceless decomposition
of the gravitational field,  following
\cite{Brown:2005aq,Kluson:2012tw}. In this procedure we introduce
new additional scalar field with additional symmetry so that the
number of physical degrees of freedom is the same. Performing
Hamiltonian analysis we also identify two second class constraints
that, together with the gauge fixing scaling symmetry allow us to
find Hamiltonian for the physical degrees of freedom, at least in
principle. These physical degrees of freedom are
 metric with unit determinant and conjugate traceless
momenta so that the number of physical degrees of freedom  is the
same as in the case of General Relativity. On the other hand there
are also important differences.  Since this theory arises from the
theory with the complicated second class constraints we find that
there is a very  complicated symplectic structure on the phase space
of the physical degrees of freedom. Secondly, even if we can claim
that these second class constraints can be solved in principle we
find that their solutions have the form of the non-local
perturbative expansions. In other words it is hard to see how such a
theory could be useful for some practical computations or even for
its path integral formulation.

However we mean that the analysis performed here suggests very
interesting direction in further research. The starting point would
be the general form of the action for the physical degrees of
freedom as was analyzed in \cite{Khoury:2013oqa} where we now
presume that the additional term in the diffeomorphism constraint
depends  either on higher order of scalar curvature as for example
$R_{ij}R^{ij}$ or it depends on $R_{ij}$ non-locally. Then we should
proceed as in \cite{Khoury:2013oqa} where we demand that the Poisson
brackets of the spatial diffeomorphism constraints close on the
constraint surface. Then from the requirement of the preservation of
these constraints during the time evolution of the system we could
determine corresponding Hamiltonian density. We hope to return to
this problem in future.

\section{Brief Review of Lagrange Multiplier
 Modified HL Gravity}\label{second} We begin this section
with the brief review of  the Lagrange multiplier modified  RFDiff
invariant HL gravity, for more detailed treatment see
\cite{Kluson:2011xx}.
 RFDiff invariant Ho\v{r}ava-Lifshitz
 gravity was introduced in
\cite{Blas:2010hb}, see also \cite{Kluson:2010na}. In
\cite{Kluson:2011xx} this action was extended by introducing
Lagrange multiplier term that ensures that the spatial curvature is
constant. Explicitly,Lagrange multiplier modified RFDiff HL gravity
has the form
\begin{equation}\label{RFDaction}
S=\frac{1}{\kappa^2} \int dt d^3\bx \sqrt{h}(\tK_{ij}
\mG^{ijkl}\tK_{kl}-\mV(h)+\mG[R]\mA) \ ,
\end{equation}
where $\mG[R]=R-\Omega$,  where $\Omega$ is constant, $\mA$ is
Lagrange multiplier that transforms as scalar
\begin{equation}
\mA'(t',\bx')= \mA(t,\bx) \
\end{equation}
under (\ref{rfdtr}). Further, $\tK_{ij}$ introduced in
(\ref{RFDaction}) is
 modified extrinsic
curvature
\begin{equation}
\tK_{ij}=\frac{1}{2}(\partial_t h_{ij} -\nabla_i N_j-\nabla_j N_i) \
\end{equation}
that differs from the standard extrinsic
curvature by absence of the lapse $N(t)$.
Further the generalized De Witt metric
$\mG^{ijkl}$  is defined as
\begin{equation}
\mG^{ijkl}=\frac{1}{2}(h^{ik}h^{jl}+ h^{il}h^{jk})-\lambda
h^{ij}h^{kl} \ ,
\end{equation}
where $\lambda$ is a real constant that in case of General
Relativity is equal to one. Finally $\mV(h)$ is a general function
of $h_{ij}$ and its covariant derivative. The analysis performed in
\cite{Kluson:2011xx} showed that this theory possesses the same
number of physical degrees of freedom as General Relativity. For
that reason we mean that this action is a good candidate for the
conformal traceless decomposition of the gravitational field and
possible identification of the physical degrees of freedom of HL
gravity.

\section{Conformal  Traceless Decomposition}

The conformal-traceless decomposition of the gravitational field was
firstly performed in \cite{York:1998hy} in its initial value problem
\footnote{For review and extensive list of references, see
\cite{Gourgoulhon:2007ue}.}. In order to implement
conformal-traceless decomposition we follow \cite{Brown:2005aq} and
define $h_{ij}$ and $\tK_{ij}$ as
\begin{equation}\label{defcon}
h_{ij}=\phi^4 g_{ij} \ , \quad
\tK_{ij}=\phi^{-2}A_{ij}+\frac{1}{3}\phi^4 g_{ij}\tau \ .
\end{equation}
 We see that this definition is redundant since the multiple of the
fields $g_{ij},\phi,A_{ij},\tau$ give the same physical metric
$h_{ij}$ and modified extrinsic curvature $\tK_{ij}$. In fact, we
see that the decomposition (\ref{defcon}) is invariant under the
conformal transformation
\begin{eqnarray}\label{gaugecon}
g'_{ij}(\bx,t)&=&\Omega^4(\bx,t)g_{ij}(\bx,t) \ , \quad
\phi'(\bx,t)=\Omega^{-1}(\bx,t)
\phi(\bx,t) \ , \nonumber \\
A'_{ij}(\bx,t)&=&\Omega^{-2}(\bx,t)A_{ij}(\bx,t), \quad
\tau'(\bx,t)=\tau(\bx,t) \ .
\nonumber \\
\end{eqnarray}
 We also see that (\ref{defcon}) is
invariant under following transformation
\begin{equation}\label{secondsym}
\tau'(\bx,t)=\tau(\bx,t)+\zeta(\bx,t) \ , \quad A'_{ij}(\bx,t)=
A_{ij}(\bx,t) -\frac{1}{3}\zeta(\bx,t) \phi^6 g_{ij}(\bx,t) \ .
\end{equation}
Clearly the gauge fixing of these symmetries we can eliminate $\tau$
and $\phi$.

 In order to perform the Hamiltonian analysis of the conformal
decomposition of the action (\ref{RFDaction}) we firstly
 rewrite the
action (\ref{RFDaction}) to its Hamiltonian form. To do this we
introduce the conjugate momenta
\begin{eqnarray}\label{defmom}
P^{ij}=\frac{\delta S}{\delta
\partial_t h_{ij}}=
\frac{1}{\kappa^2}\sqrt{h}\mG^{ijkl}\tK_{kl} \ , \quad
P_i=\frac{\delta S}{\delta \partial_t N^i}=
0 \ , \quad P_{\mA}=\frac{\delta S}{\delta \partial_t\mA}\approx 0 \ . \nonumber \\
\end{eqnarray}
Then we easily determine corresponding
 Hamiltonian
\begin{equation}
H=\int d^3\bx (\partial_t h_{ij}P^{ij}-\mL)= \int d^3\bx
(\mH'_T+N^i\mH'_i) \ ,
\end{equation}
where
\begin{equation}
\mH'_T=\frac{\kappa^2}{\sqrt{h}}
P^{ij}\mG_{ijkl}P^{kl}+\sqrt{g}\mV(h)-\sqrt{h}\mA\mG(R) \ ,
 \quad
\mH'_i=-2h_{ij}\nabla_k P^{jk} \ .
\end{equation}
Using the Hamiltonian and the corresponding canonical variables we
write the action (\ref{RFDaction}) as
\begin{equation} S=\int dt L=
\int dt d^3 \bx (P^{ij}\partial_t h_{ij}-\mH)= \int dt d^3\bx
(P^{ij}\partial_t h_{ij}-N \mH'_T- N^i\mH'_i) \ .
\end{equation}
Then we insert the decomposition (\ref{defcon}) into the definition
of the canonical momenta $P^{ij}$
\begin{equation}\label{Pabh}
P^{ij}=\frac{1}{\kappa^2} \sqrt{g}(\phi^{-4}\tmG^{ijkl}A_{kl}
+\frac{1}{3}\phi^2\tau \tmG^{ijkl}g_{kl}) \ ,
\end{equation}
where the metric $\tmG^{ijkl}$ is defined as
\begin{equation}
\tmG^{ijkl}=\frac{1}{2}(g^{ik}g^{jl}+g^{il}g^{jk})-\lambda
g^{ij}g^{kl} \ , \quad \mG^{ijkl}=\phi^{-8}\tmG^{ijkl} \ .
\end{equation}
Note that $\tmG^{ijkl}$ has the inverse
\begin{equation}
\tmG_{ijkl}=\frac{1}{2}(g_{ik}g_{jl}+
g_{il}g_{jk})-\frac{\lambda}{3\lambda-1} g_{ij}g_{kl} \ , \quad
\tmG_{ijkl}= \phi^8 \mG_{ijkl} \ .
\end{equation}
Using (\ref{Pabh}) and (\ref{defcon}) we rewrite $P^{ij}\partial_t
h_{ij}$ into the form
\begin{eqnarray}
P^{ij}\partial_t h_{ij}&=&
\left(\frac{1}{\kappa^2}\sqrt{g}\tmG^{ijkl}A_{kl} +
\frac{\sqrt{g}}{3\kappa^2} \phi^6(1-3\lambda)\tau
g^{ij}\right)\partial_t g_{ji}+ \nonumber \\
&+& \left(\frac{4}{\kappa^2}\sqrt{g}\phi^{-1}
A_{kl}g^{kl}(1-3\lambda)+\frac{4\sqrt{g}}{\kappa^2}
(1-3\lambda)\phi^5\tau\right)\partial_t\phi
\ .  \nonumber \\
\end{eqnarray}
We see that it is natural to identify the expression in the
parenthesis with  momentum $\pi^{ij}$ conjugate to $g_{ij}$  and
$p_\phi$ conjugate to $\phi$ respectively
\begin{eqnarray}\label{canmom}
\pi^{ij}&=&\frac{1}{\kappa^2}
\sqrt{g}\tmG^{ijkl}A_{kl}+\frac{\sqrt{g}}{3\kappa^2}(1-3\lambda)
\phi^6\tau g^{ij} \ , \nonumber \\
p_\phi&=&\frac{4}{\kappa^2} \sqrt{g}\phi^{-1}
A_{ij}g^{ji}(1-3\lambda)+ \frac{4\sqrt{g}}{\kappa^2}
(1-3\lambda)\phi^5 \tau \ . \nonumber
\\
\end{eqnarray}
Then using (\ref{canmom}) we obtain  following primary constraint
\begin{equation}
\Sigma_D: p_\phi \phi-4\pi^{ij}g_{ji}=0 \ .
\end{equation}
As we will see below this is the constraint that generates conformal
transformation of the dynamical fields.
 Further,  using
(\ref{canmom}) we find the relation between $P^{ij}$ and $\pi^{ij}$
in the form
\begin{equation}
P^{ij}=\phi^{-4}\pi^{ij} \ .
\end{equation}
Then we find that the kinetic term in  $\mH_T$ takes the form
\begin{equation} \frac{\kappa^2}{\sqrt{h}}
P^{ij}\mG_{ijkl}P^{kl}= \frac{\kappa^2\phi^{-6}}{\sqrt{g}}
\pi^{ij}\tmG_{ijkl}\pi^{kl} \ .
\end{equation}
 As the next step we introduce the
decomposition (\ref{defcon}) into the contribution $ \int d^3\bx N^i
\mH'_i$. Using the  relation between Levi-Civita connections
 evaluated with the metric components $h_{ij}$ and
$g_{ij}$
\begin{equation}
\Gamma_{ij}^k(h)=\Gamma_{ij}^k(g)+ 2\frac{1}{\phi} (\partial_i\phi
\delta^k_j+\partial_j \phi\delta_i^k-\partial_l\phi g^{kl}g_{ij}) \
\end{equation}
and also if we define
 $n_i$ through the relation
 $N_i=\phi^4n_i$
we obtain
\begin{eqnarray}
\int d^3\bx N^i\mH'_i&=&
\int d^3\bx n^i \mH''_i \ , \nonumber \\
\end{eqnarray}
where
\begin{equation}
 \mH''_i=-2g_{ik}D_j \pi^{jk}+
4\phi^{-1}\partial_i \phi g_{kl}\pi^{kl} \ ,
 \nonumber \\
\end{equation}
where the covariant derivative $D_i$ is defined using the
Levi-Civita connection $\Gamma^k_{ij}(g)$.
Observe that with the help of the constraint $\Sigma_D $
 we can write the constraint $\mH''_i$ as
\begin{eqnarray}
\mH''_i=
 -2g_{ik}D_j \pi^{jk}+ \partial_i\phi p_\phi-
4\phi^{-1}\partial_i\phi \Sigma_D\equiv  \hat{\mH}_i -
4\phi^{-1}\partial_i\phi \Sigma_D
\end{eqnarray}
so that we see that it is natural to
 identify $\hat{\mH}_i$ as an independent constraint. In
 fact, we will see that the smeared form of this constraint
generates the spatial diffeomorphism.

Finally we should proceed to the analysis of the spatial curvature
and generally the whole potential term $\mV$. Note that this is the
function of the covariant derivative, $R$ and $R_{ij}$. Using the
following formulas
\begin{eqnarray}
R_{ij}[h]&=& R_{ij}[g]+\frac{6}{\phi^2}D_i\phi D_j\phi-
\frac{2}{\phi} D_iD_j\phi-2\frac{g_{ij}}{\phi}D_k[g^{kl}D_l\phi]
-\frac{2}{\phi^2}
g_{ij}D_k\phi g^{kl}D_l\phi \ , \nonumber \\
R[h]&=&\phi^{-4}[ R[g] -\frac{8}{\phi}
 g^{ij}D_iD_j\phi
] \ , \nonumber \\
\end{eqnarray}
Then using also the relation between Levi-Civita connections
evaluated on $h$ and $g$ we find that the potential term is
generally function of $\phi$ and $g$ whose explicit form is not
needed here.  As a result we find the action in the form
\begin{eqnarray}\label{actiontraceless}
S&=&\int dt d^3\bx(\pi^{ij}\partial_t g_{ij}+ p_\phi\partial_t\phi
-n^i\hat{\mH}_i-\frac{\kappa^2\phi^{-6}}{\sqrt{g}}
\pi^{ij}\tmG_{ijkl} \pi^{kl}-\sqrt{g}\phi^6\mV(\phi,h)+\nonumber \\
&+&\sqrt{g}\phi^6 \mA
 \mG(\phi^{-4}R[g]-\frac{8}{\phi^5}g^{ij}D_iD_j\phi)
 -\lambda
\Sigma_D) \ ,
\nonumber \\
\end{eqnarray}
where we included the primary constraint $\Sigma_D$ multiplied by
the Lagrange multiplier $\lambda$.

Now we can proceed to the Hamiltonian analysis of the conformal
decomposition of the gravitational field given  by the  action
(\ref{actiontraceless}).
 Clearly we have following primary
constraints
\begin{equation}
 \pi_i\approx 0 \ , \pi_{\mA}\approx 0 \ ,
\quad \Sigma_D\approx 0 \ ,
\end{equation}
where $\pi_i$ and $\pi_{\mA}$ are momenta conjugate to $n^i$ and
$\mA$
 with following
non-zero Poisson brackets
\begin{equation}
\pb{n^i(\bx),\pi_j(\by)}= \delta^i_j\delta(\bx-\by) \ , \quad
\pb{\mA(\bx),\pi_{\mA}(\by)}=\delta(\bx-\by) \ .
\end{equation}
Further, the preservation of the primary constraints $\pi_i$ and
$\pi_{\mA}$  implies following secondary ones
\begin{equation}
\hat{\mH}_i\approx 0 \ , \Phi_1 \equiv
\frac{1}{\kappa^2}\sqrt{g}\phi^6 \mG\approx 0 \ .
\end{equation}
Now we should analyze the requirement of the preservation of the
primary constraint $\Sigma_D$ during the time evolution of the
system. First of all the explicit calculations give
\begin{eqnarray}\label{SigmaDg}
\pb{\Sigma_D(\bx),g_{ij}(\by)}&=&
4g_{ij}(\bx)\delta(\bx-\by) \ , \nonumber \\
\pb{\Sigma_D(\bx),\pi^{ij}(\by)}&=& -4\pi^{ij}(\bx)\delta(\bx-\by) \
, \nonumber
\\
\pb{\Sigma_D(\bx),\phi(\by)}&=& -\phi(\bx)\delta(\bx-\by) \ ,
\nonumber
\\
\pb{\Sigma_D(\bx),p_\phi(\by)}&=& \phi(\bx)\delta(\bx-\by) \
\nonumber
\\
\end{eqnarray}
using the canonical Poisson brackets
\begin{equation}
\pb{g_{ij}(\bx),\pi^{kl}(\by)}= \frac{1}{2} (\delta_i^k\delta_j^l+
\delta_i^l\delta_j^k)\delta(\bx-\by) \ , \quad
\pb{\phi(\bx),p_\phi(\by)}=\delta(\bx-\by) \ .
\end{equation}
 It turns out that it is useful
to introduce the smeared forms of the constraints
$\hat{\mH}_i,\Sigma_D$
\begin{equation}
\bT_S(N^i)= \int d^3\bx N^i \hat{\mH}_i \ , \quad \bD(M)=\int d^3\bx
M \Sigma_D \ ,
\end{equation}
where $N^i$ and $M$ are smooth functions on $\mathbf{R}^3$.  Then
using (\ref{SigmaDg}) and also
\begin{eqnarray}
\pb{\Sigma_D(\bx),\Gamma_{ij}^k(\by)}=2
\delta_j^k\partial_{y^i}\delta(\bx-\by)+2\delta_i^k\partial_{y^j}
\delta(\bx-\by)-2g^{kl}(\by)\partial_{y^l}\delta(\bx-\by)g_{ij}
(\by) \
\nonumber \\
\end{eqnarray}
we easily find that
\begin{equation}\label{bDmHT}
\pb{\bD(M),\mH'_T(\by)}=0 \ ,
\end{equation}
where
\begin{equation}
\mH'_T=\frac{\kappa^2\phi^{-6}}{\sqrt{g}} \pi^{ij}\tmG_{ijkl}
\pi^{kl}+\sqrt{g}\phi^6\mV(\phi,g) \ .
\end{equation}
 To proceed further we use following Poisson brackets
\begin{eqnarray}\label{pbbtSgp}
\pb{\bT_S(N^i),g_{ij}(\bx)}&=&-N^k\partial_k g_{ij}(\bx)-
\partial_i N^kg_{kj}(\bx)-g_{ik}\partial_j N^k(\bx)  \ , \nonumber \\
\pb{\bT_S(N^i),\pi^{ij}(\bx)}&=&-\partial_k (N^k\pi^{ij})(\bx) +
\partial_k N^i\pi^{kj}(\bx)+\pi^{ik}\partial_k N^j(\bx) \ ,  \nonumber \\
\pb{\bT_S(N^i),\phi(\bx)}&=&-N^i\partial_i\phi(\bx) \ , \nonumber \\
\pb{\bT_S(N^i),p_\phi(\bx)}&=&-\partial_i (N^i p_\phi)(\bx) \
\nonumber \\
\end{eqnarray}
and hence it is easy to see that
\begin{equation}
\pb{\bT_S(N^i),\Sigma_D(\bx)}=-N^i\partial_i \Sigma_D(\bx)-
\partial_i N^i\Sigma_D(\bx) \
\end{equation}
that together with (\ref{bDmHT}) implies that $\Sigma_D \approx 0$
is the first class constraint.

Now we proceed to the analysis of the preservation of the secondary
constraints $ \hat{\mH}_i\approx 0$ and $\Phi_1\approx 0$. Note that
the total Hamiltonian takes the form
\begin{eqnarray}
H_T=\int d^3\bx (\mH_T'
+\lambda \Sigma_D
+n^i\hat{\mH}_i+\gamma p_{\mA}+\Gamma_{I}\Phi_1 ) \ ,  \nonumber \\
\end{eqnarray}
where $\gamma$ is the Lagrange multiplier corresponding to the
constraint $p_{\mA}$ while $\Gamma_{I}$ is the Lagrange multiplier
corresponding to the constraint $\Phi_1\approx 0$.

  In case
of $\hat{\mH}_i$ we find following Poisson brackets
\begin{equation}
\pb{\hat{\mH}_i(\bx),\hat{\mH}_j(\by)}=
\hat{\mH}_j(\bx)\frac{\partial} {\partial
x^i}\delta(\bx-\by)-\hat{\mH}_i(\by)\frac{\partial}{\partial y^j}
\delta(\bx-\by) \
\end{equation}
which implies that the smeared form of the diffeomorphism
constraints takes the familiar form
\begin{equation}
\pb{\bT_S(N^i),\bT_S(M^j)}= \bT_S(N^j\partial_j M^i- M^j\partial_j
N^i) \ .
\end{equation}
Further using (\ref{pbbtSgp}) we easily find
\begin{eqnarray}
\pb{\bT_S(N^i),\mH'_T(\bx)}= -\partial_i N^i\mH'_T(\bx) -
N^i\partial_i \mH'_T(\bx) \ , \nonumber \\
\pb{\bT_S(N^i),\Phi_1(\bx)}=-\partial_i N^i\Phi_1(\bx)-
N^i\partial_i\Phi_1(\bx) \  \nonumber \\
\end{eqnarray}
that implies that $\hat{\mH}_i$ are the first class constraints
that are preserved during the time evolution of the system.
Finally we analyze the time evolution of the constraint
 $\Phi_1\approx
0$
. Using following formulas
\begin{eqnarray}
\pb{R(\bx),\pi^{ij}(\by)}&=& -R^{ij}(\bx)\delta(\bx-\by)+D^i D^j
\delta(\bx-\by)-g^{ij} D_k D^k\delta(\bx-\by) \ ,
\nonumber \\
\pb{\Gamma_{ij}^k(\bx),\pi^{mn}(\by)}&=& \frac{1}{4}g^{kp} [
D_i\delta (\delta_j^m\delta_p^n+
\delta_j^n\delta_p^m)\delta(\bx-\by)+\nonumber \\
 &+& D_j (\delta_p^m\delta_i^n+\delta_p^n\delta_i^m)\delta(\bx-\by)-
 D_p (\delta_i^m\delta_j^n+\delta_i^n\delta_j^m)\delta(\bx-\by)]
 \nonumber \\
\end{eqnarray}
we find that the time derivative of $\Phi_1$ is equal to
\begin{eqnarray}
\partial_t\Phi_1&=&\pb{\Phi_1,H_T}\approx
\nonumber \\
&-&\frac{2\kappa^2 }{\phi^4\sqrt{g}} (R_{ij}\pi^{ij}-
\frac{\lambda}{3\lambda-1}R\pi)+\nonumber \\
&+&\frac{2\kappa^2\phi^2}{\sqrt{g}}D_kD_l[\phi^{-6}\pi^{kl}] +
\frac{2\kappa^2\phi^2}{\sqrt{g}}\frac{(1-\lambda)}{3\lambda-1}D_k
D^k[\phi^{-6}\pi]
-\nonumber \\
 &-&16\frac{\kappa^2}{\phi^5\sqrt{g}}
(\pi^{ij}-\frac{\lambda}{3\lambda-1}g^{ij}\pi)D_iD_j\phi+\nonumber
\\
&+&\frac{16 \phi\kappa^2}{\sqrt{g}}D_i\phi D_j[\phi^{-6}\pi^{ij}]-
\frac{8\kappa^2\phi}{\sqrt{g}}\frac{2\lambda-1}{3\lambda-1}D_i\phi
g^{ij} D_j[\phi^{-6}\pi]\equiv \Phi_2 \ , \nonumber \\
\end{eqnarray}
where $\Phi_2$  is an  additional constraint that has to be imposed
on the system. Following
 \cite{Henneaux:1992ig,Govaerts:2002fq,Govaerts:1991gd}
 we include the constraint $\Phi_2$ into
the  definition of the total Hamiltonian that now has the form
\begin{eqnarray}
H_T&=&\int d^3\bx (\mH_T'-\sqrt{g}\phi^6\mA (\phi^{-4}R[g]-
8\phi^{-5}g^{ij}D_iD_j\phi-\Omega)+\lambda \Sigma_D+\nonumber
\\
&+&n^i\hat{\mH}_i+\gamma p_{\mA}+\Gamma_{I}\Phi_1 +\Gamma_{II}\Phi_2
 ) \ .  \nonumber \\
\end{eqnarray}
Now we should again check the stability of all constraints. It is
easy to see that the primary constraints together with $\bT_S(N^i)$
are preserved while the   time  evolution of the constraint
$\Phi_1\approx 0$ is equal to
\begin{eqnarray}\label{parttPhi2}
\partial_t \Phi_1&=&\pb{\Phi_1,H_T}
\approx  \int d^3\bx \left(\Gamma^{II}(\bx)
\pb{\Phi_1,\Phi_{2}(\bx)}\right)
\approx \nonumber \\
&\approx& \int d^3\bx \Gamma^{II}(\bx)\pb{\Phi_1,\Phi_{2}(\bx)}=0 \
.
\nonumber \\
\end{eqnarray}
As follows from the explicit form of the constraints $\Phi_{1,2}$ we
have
\begin{eqnarray}\label{DBphi12}
\pb{\Phi_1(\bx),\Phi_{2}(\by)}\neq 0 \ .
\end{eqnarray}
Then  we find that the equation (\ref{parttPhi2}) gives
$\Gamma^{II}=0$. In the same way the requirement of the preservation
of the constraint $\Phi_{2}$ implies
\begin{eqnarray}\label{partPhi2}
\partial_t\Phi_{2}\approx
\int d^D\bx (\pb{\Phi_{2},\mH_T(\bx)} +
\Gamma^I(\bx)\pb{\Phi_{2},\Phi_{1}(\bx)})=0 \ .
\nonumber \\
\end{eqnarray}
Using the fact that $\pb{\Phi_{2},\mH_T(\bx)}\neq 0$ and also the
equation (\ref{DBphi12}) we see that (\ref{partPhi2}) can be solved
for $\Gamma^I$. In fact,  (\ref{DBphi12}) shows that
 $\Phi_1$ and
$\Phi_{2}$ are the second class constraints. We also see from the
previous analysis that no additional constraints have to be imposed
on the system.

In order to find the action for the physical degrees of freedom we
have to finally fix the gauge symmetry generated by $\Sigma_D$. To
do this we introduce the gauge fixing function
\begin{equation}
\Phi_{G.F.}=\sqrt{g}-1 \ .
\end{equation}
It is easy to see that there is non-zero Poisson bracket between
$\Phi_{G.F.}$ and $\Sigma_D$ so that they are the second class
constraints. In summary we have following collection of the second
class constraints
\begin{equation}
\Phi_1=0  \ , \Phi_2=0  \ , \Sigma_D=0 \ , \Phi_{G.F.}=0 \ .
\end{equation}
The goal is to eliminate some degrees of freedom from these
constraints, at least in principle. In fact, from $\Phi_1$, which
is version of Lichnerowitz-York equation
 \cite{York:1971hw}, we
express $\phi$ as
\begin{equation}
\phi=\frac{1}{8}\nabla^{-1}(\phi R[g]-\phi^5\Omega) \ .
\end{equation}
where $\nabla^{-1}$ is inverse operator to $g^{ij}D_iD_j$. We can
solve the equation above perturbatively around some constant
$\phi_0$. Further, from $\Sigma_D$ we express $p_\phi$. Finally,
$\Phi_{G.F.}$ reduces number of degrees of freedom in $g$ to be
equal to five and from $\Phi_2$ we express $\pi$ as the function of
remaining degrees of freedom. In summary, the physical degrees of
freedom of Lagrange multiplier modified HL gravity are
\begin{equation}
g_{ij}, \quad \sqrt{g}=1 \ , \quad \tilde{\pi}^{ij} \ ,
g_{\ij}\tilde{\pi}^{ji}=0 \ .
\end{equation}
Note that there are three first class constraints $\hat{\mH}_i$ so
that by gauge fixing these constraints we should eliminate remaining
three degrees of freedom in $g_{ij}$. In other words the physical
content of given theory is the same as in General Relativity.

From the previous analysis we see that $\phi$  is a non-local
function of $R$. The same situation also occurs in case of $\pi$ so
that in principle $\hat{\mH}_i$ has the form
\begin{equation}
\hat{\mH}_i=-2g_{ik}\nabla_j
\tilde{\pi}^{kj}+p_\phi(\nabla^{-1}f(g,\pi))\nabla_i
\phi(\nabla^{-1}R(g))-\frac{1}{3}\nabla_i
\pi(\nabla^{-1},\tilde{\pi},g) \ .
\end{equation}
 By definition they are the first class constraints which however
 depends non-locally on the canonical variables.
 We would like to  stress
 that this result can be considered as the generalization of
  the  analysis performed
in \cite{Khoury:2013oqa,Khoury:2011ay} to the case of HL gravity. In
more details,   papers cited above  were devoted to the construction
of the action for the physical modes of the gravitational fields
only that are the metric that obeys the condition $\sqrt{g}=1$ and
the conjugate momenta $\tilde{\pi}^{ij}$ that are traceless. In
order to have the right number of physical degrees of freedom this
action should be invariant under  spatial diffeomorphism.
Explicitly, the action studied there  has the form
\begin{equation}
S=\int dt d^3\bx
(\dot{g}_{ij}\tilde{\pi}^{ij}-\pi_H-N^i\tilde{\mH}_i) \ ,
\end{equation}
where
\begin{equation}\label{KhourimH}
\tilde{\mH}_i=-2g_{ij}\nabla_k \tilde{\pi}^{jk}-\nabla_i \pi_K \ ,
\end{equation}
where $\pi_K$ is arbitrary function that has to be determined in
such a way that $\tilde{\mH}_i$ are the first class constraints.
However when we restrict to the case when this function depends on
the partial derivatives of $g_{ij}$ through the scalar curvature $R$
we find that the functions $\pi_N$ and $\pi_K$ are uniquely
determined and leads to the General Relativity action. Note also
that the symplectic structure used in given paper has the standard
form
\begin{equation}\label{stsymp}
\pb{g_{ij},\tilde{\pi}^{kl}}=\frac{1}{2}(\delta_i^k
\delta_j^l+\delta_i^l \delta_j^k)-\frac{1}{3}g_{ij}g^{kl} \ .
\end{equation}
In case of Lagrange multiplier modified HL gravity the situation is
more involved. The symplectic structure is determined by the Poisson
brackets of the second class constraints $\Phi_1,\Phi_{2}$ which is
rather complicated. More precisely,  let us denote all second class
constraints as $\Phi_A=(\Sigma_D,\Phi_{G.F.},\Phi_1,\Phi_2)$. Then
the Poisson bracket between the constraints $\Phi_A$  can be written
as
\begin{equation}
\pb{\Phi_A(\bx),\Phi_B(\by)}= \triangle_{AB}(\bx,\by) \ ,
\end{equation}
where  the matrix $\triangle_{AB}$ has following structure
\begin{equation}\label{triangleAB}
\triangle_{AB}(\bx,\by)= \left(\begin{array}{cccc}
0 & * & 0 & 0   \\
 {*} & 0 & 0 & * \\
0 & 0 & 0 & * \\
0 & * & * & * \\
\end{array}\right) \ ,
\end{equation}
where $*$ denotes non-zero elements  that depend on the phase space
variables and their derivatives.
Now it is easy to see that the Dirac brackets between the canonical
variables  that are defined as
\begin{eqnarray}
& &\pb{g_{ij}(\bx),g_{kl}(\by)}_D= -\int d\bz d\bz'
\pb{g_{ij}(\bx),\Phi_A(\bz)} (\triangle^{-1})^{AB}(\bz,\bz')
\pb{\Phi_B(\bz'),g_{kl}(\by)}\ ,
\nonumber \\
& &\pb{\pi^{ij}(\bx),\pi^{kl}(\by)}_D= -\int d\bz d\bz'
\pb{\pi^{ij}(\bx),\Phi_A(\bz)} (\triangle^{-1})^{AB}(\bz,\bz')
\pb{\Phi_B(\bz'),\pi^{kl}(\by)}
\ ,
\nonumber \\
& &\pb{g_{ij}(\bx),\pi^{kl}(\by)}_D= \pb{g_{ij}(\bx),\pi^{kl}(\by)}
-\nonumber \\
&-&\int d\bz d\bz' \pb{g_{ij}(\bx),\Phi_A(\bz)}
(\triangle^{-1})^{AB}(\bz,\bz') \pb{\Phi_B(\bz'),\pi^{kl}(\by)} \
\nonumber \\
\end{eqnarray}
 depend on the phase-space variables.
 Secondly, the resulting
action is non-local due the presence of the inverse operator
$\nabla^{-1}$.

Even if the action for the physical degrees of freedom of Lagrange
multiplier modified HL gravity is rather involved we mean that the
results derived here should be considered as the starting point for
further research of HL gravity when  we generalize the analysis
performed in \cite{Khoury:2013oqa,Khoury:2011ay} in several
different ways. We could start with the action for the physical
modes with the symplectic structure given by the equation
(\ref{stsymp}) and presume that $\pi_K$ depends on $R$ non-locally
and try to determine the original action. Another possibility is to
consider the case when $\pi_K$ in (\ref{KhourimH}) depends on
$R_{ij}R^{ij}$ and covariant derivatives of $R_{ij}$. We hope to
return to these problems in future.
\\
\noindent {\bf Acknowledgement}

 This work  was supported by the
Grant Agency of the Czech Republic under the grant P201/12/G028.

\end{document}